\documentclass[a4paper,fleqn,usenatbib]{mnras}

\usepackage{mathptmx}

\usepackage[T1]{fontenc}
\usepackage{ae,aecompl}

\usepackage{graphicx}	
\usepackage{amsmath}	
\usepackage{amssymb}	
\usepackage{pdflscape}  

\title[Testing EEP with short GRBs]{Testing Einstein's Equivalence Principle with Short Gamma-ray Bursts}

\author[Y. Sang, H.-N. Lin and Z. Chang]{
Yu Sang,$^{1}$\thanks{E-mail: sangyu@ihep.ac.cn}
Hai-Nan Lin,$^{2}$
and Zhe Chang$^{1}$
\\
$^{1}$Institute of High Energy Physics, Chinese Academy of Sciences, Beijing 100049, China\\
$^{2}$Department of Physics, Chongqing University, Chongqing 401331, China\\
}

\date{Accepted XXX. Received YYY; in original form ZZZ}

\pubyear{2016}

\begin{document}
\label{firstpage}
\pagerange{\pageref{firstpage}--\pageref{lastpage}}
\maketitle

\begin{abstract}
Einstein's equivalence principle (EEP) can be tested by the time delay between photons with different energies passing through a gravitational field. As one of the most energetic explosions in the Universe, gamma-ray bursts (GRBs) provide an effective tool to test the accuracy of EEP. In this paper, we use the continuous spectra of 20 short GRBs detected by the Swift/BAT to test the validity of EEP. Taking the duration of GRBs as the upper limit of the time delay induced by EEP violation (assuming that the high energy photons arrive later than the low energy photons), the difference of the parameterized post-Newtonian parameter is constrained with high accuracy. The strictest constraint, $|\gamma(150~{\rm keV})-\gamma(15~{\rm keV})|<5.59\times 10^{-10}$ from GRB 150101B, is about $1\sim 2$ orders of magnitude tighter than previous constraints. Moreover, our result is more statistically significant than previous results because we use the continuous spectra instead of isolated photons.
\end{abstract}




\section{Introduction}

Einstein's equivalence principle (EEP) is one of the foundations of general relativity, and it is a more rigorous concept compared to the weak equivalence principle. One presentation of EEP is that the trajectory of a free test body travelling in a gravitational field is independent of its internal structure and composition \citep{Will:2014}. As massless and uncharged particles, photons can be used to test EEP in the parameterized post-Newtonian (PPN) formalism where the spacetime metric is parameterized by the coefficients (i.e., the PPN parameters) that appear in front of the metric potentials \citep{Will:2014}. If EEP is valid, then $\gamma$ is independent of photon energy. For example, $\gamma=1$ in general relativity. Here, $\gamma$ is one of the PPN parameters, and it describes how much space curvature is produced by unit rest mass. One way to test the validity of EEP is to constrain the absolute value of $\gamma$. Many experiments have been realized to provide precise constraint on it. The deflection of light has been measured using very long baseline radio interferometry, and $(\gamma-1)$ is constrained to be at the order of $10^{-4}$ \citep{Lebach:1995,Fomalont:2009,Shapiro:2004,Lambert:2009,Lambert:2011}. \citet{Bertotti:2003} measured the time delay of radio photons on round trip to the Cassini spacecraft and arrived at the constraint $\gamma-1=(2.1\pm2.3)\times10^{-5}$.

An alternative way to test the validity of EEP is to compare the values of $\gamma$ for photons with different energies \citep{Wei:2015,Gao:2015}. There is a small time delay ($\delta t$) for a photon to pass through a given distance if there is a gravitational field along the light path \citep{Shapiro:1964,Krauss:1988,Longo:1988}. If EEP is valid, then $\gamma$ is a constant, and two photons with different energies will suffer an equal $\delta t$. On the other hand, if EEP is invalid, two photons with different energies will suffer an unequal $\delta t$, thus producing an observable time delay. The relative time delay between two photons of different energies emitted simultaneously and passing through the same gravitational field constrains the difference of $\gamma$. This provides a useful way to test the validity of EEP. Based on this method, several studies have provided strict constraints on $\Delta\gamma$ \citep{Sivaram:1999,Gao:2015,Wei:2015,Wei:2016A,Nusser:2016,Wu:2016,Zhang:2016obv,Wang:2016lne,Wei:2016ygk,Tingay:2016tgf}. For example, \citet{Gao:2015} proposed that the time delay between correlated photons from gamma-ray bursts (GRBs) can be used to test EEP. They considered the gravitational field of the Milky Way and constrained the differences of $\gamma$ of photons with energies between eV and MeV or between MeV and GeV to the order of $10^{-7}$. Fast radio bursts (FRBs) were also used to test EEP. \citet{Wei:2015} used FRB/GRB 100704A system to give a limit as $|\gamma(1.23\textrm{ GHz})-\gamma(1.45\textrm{ GHz})|<4.36\times10^{-9}$, expanding the energy range of testing EEP out to the radio band. However, \citet{Palaniswamy2014} showed that FRB/GRB 100704A is unlikely to be an astrophysical event and therefore not appropriate for any calculation of EEP violation. Using the first localized burst FRB 150418, \citet{Tingay:2016tgf} obtained the best current limit, i.e., $\Delta\gamma<1-2\times10^{-9}$. \citet{Wei:2016A} proposed that the TeV blazars can also provide an excellent tool to constrain EEP, and extending the tested energy range out to TeV energy. It is worth mentioning that the time delay of different kinds of particles was also used to test EEP. For example, the time delays of photons and neutrinos from supernova 1987A in the Large Magellanic Cloud are proven to be equal to a very high accuracy \citep{Krauss:1988,Longo:1988}.

GRBs are one of the most energetic explosions in the Universe \citep{Piran:1999,Meszaros:2006rc,Kumar:2014upa}. They can be detectable up to redshift $z\sim 10$. The spectra of GRBs peak at the energy range of keV to MeV, and photons with energies as high as tens GeV can also be observed in some brightest GRBs \citep{Abdo:2009,Abdo:2009pg,Ackermann:2010us,Ackermann:2011bu}. As cosmological transients, GRBs are widely used to test the fundamental physics. For example, GRBs provide an effective way to probe the Lorentz invariance violation (LIV) effect \citep{Chang:2016,Chang:2012,Zhang:2014wpb,Amelino-Camelia:1998,Ellis:2006,Ellis:2008,Jacob:2008,Vasileiou:2015}. Constraints on LIV and EEP are both based on the time delays of photons with different energies. Some studies have used the time delays between one or some isolated high energy photons and low energy photons to give strict constraints on LIV or EEP. However, the results are not statistically significant, because the number of high energy photons is too small, and we cannot know whether high energy photons and low energy photons are emitted simultaneously. In our previous work \citep{Chang:2016}, we considered the continuous spectra of Swift/BAT short GRBs in the energy band $15-150$ keV, and used the duration of short GRBs as the upper limit to constrain LIV.

In this paper, we use 20 short GRBs with well measured redshifts from Swift data archive\footnote{\url{http://swift.gsfc.nasa.gov/archive/grb_table/}} to test the validity of EEP. The duration of GRBs is used as the conservative limit of time delay induced by EEP violation. The time delay due to EEP violation couldn't be long than the duration of GRBs. The duration corresponds to the spectra in the full Swift/BAT energy band 15 -- 150 keV. There are several reasons why we use the duration of GRBs in energy band 15 -- 150 keV. First, the energy spectrum is continuous at the keV order, and a large amount of photon events can be recorded. Thus, using keV photons to constrain EEP is more statistically reliable. Second, we are not sure whether photons in different energy bands (e.g., GeV photons and MeV photons) are emitted simultaneously. However, it is very likely that photons in the same energy bands (e.g., keV photons) are emitted simultaneously, or the intrinsic time delays between them are very small. Finally, the LIV-induced time delay for keV photons is negligible.

The rest of this paper is arranged as follows. In Section II, we shortly describe the method of testing EEP. In Section III, we use 20 short GRBs from the Swift data archive to give strict constraints on EEP. Finally, discussions and conclusions are given in Section IV.

\section{Methodology}

Suppose two photons with different energies emit from a cosmological source. The observed time delay between them consists of five parts,
\begin{equation}\label{observed time delay}
\Delta t_\textrm{obs}=\Delta t_\textrm{int}+\Delta t_\textrm{LIV}+\Delta t_\textrm{spe}+\Delta t_\textrm{DM}+\Delta t_\textrm{gra},
\end{equation}
where $\Delta t_\textrm{int}$ is the intrinsic time delay which depends on the emission mechanism of the source, $\Delta t_\textrm{LIV}$ is the time delay induced by the LIV effect, $\Delta t_\textrm{spe}$ is the time delay caused by the special-relativistic effect when the rest mass of photon is non-zero, $\Delta t_\textrm{DM}$ is the time delay caused by the dispersion of photons with the free electrons along the line-of-sight, and $\Delta t_\textrm{gra}$ is the Shapiro  time delay produced when two photons with different energies traverse the gravitational field.

Since the emission mechanism of GRBs is not clearly known, $\Delta t_{\rm int}$ is highly uncertain. However, spectral observations show that the high energy (GeV) photons, at least in some GRBs, have spectral lag relative to low energy (MeV) photons \citep{Abdo:2009,Abdo:2009pg,Ackermann:2010us,Ackermann:2011bu}. Therefore, it is very likely that high energy photons are emitted later than low energy photons, thus $\Delta t_{\rm int}>0$. $\Delta t_\textrm{LIV}$ is very small even for GeV photons \citep{Ellis:2006,Ellis:2008,Jacob:2008,Vasileiou:2015}, and for keV photons it is much smaller. Hence $\Delta t_\textrm{LIV}$ is negligible. $\Delta t_\textrm{spe}$ is proportional to the square of the photon rest mass \citep{Gao:2015}, while the latter has proved to be very close to zero \citep{Tu:2005ge,Wu:2016brq}. Therefore, $\Delta t_\textrm{spe}$ is also negligible. $\Delta t_\textrm{DM}$ is inversely proportional to the square of photon energy \citep{Ioka:2003fr}, and it is only important in the radio band. For $\gamma$-ray photons in the keV energy band, $\Delta t_\textrm{DM}$ is negligible. The relative Shapiro time delay $\Delta t_\textrm{gra}$ for two photons with energy $E_1$ and $E_2$ travelling the gravitational field $U\left(r\right)$ can be written as \citep{Shapiro:1964uw}
\begin{equation}\label{gravitational potential time delay}
\Delta t_\textrm{gra}=\frac{\gamma_1-\gamma_2}{c^3}
\int^{r_\textrm{e}}_{r_\textrm{o}}U\left(r\right)dr,
\end{equation}
where $r_\textrm{e}$ and $r_\textrm{o}$ are the locations of source and observer, respectively. In summary, the middle three terms on the right-hand-side of equation~(\ref{observed time delay}) are negligible, and the contribution to the observed time delay mainly comes from $\Delta t_\textrm{int}$ and $\Delta t_\textrm{gra}$, i.e.,
\begin{equation}\label{reduced observed time delay}
\Delta t_\textrm{obs}=\Delta t_\textrm{int}+\Delta t_\textrm{gra}.
\end{equation}
Under the assumption that $\Delta t_\textrm{int}>0$, we obtain the following inequality,
\begin{equation}\label{PPN inequation}
\Delta t_\textrm{obs}>\frac{\gamma_1-\gamma_2}{c^3}
\int^{r_\textrm{e}}_{r_\textrm{o}}U\left(r\right)dr.
\end{equation}

In general, for a cosmological source, there are three parts contributing to $U\left(r\right)$, i.e., $U\left(r\right)=U_\textrm{MW}\left(r\right)+U_\textrm{IG}\left(r\right)+U_\textrm{host}\left(r\right)$, including the gravitational potential of the Milky Way, the intergalactic background between host galaxy and the Milky Way, and the host galaxy of the source. The potential models for $U_\textrm{IG}\left(r\right)$ and $U_\textrm{host}\left(r\right)$ are unknown, but it is plausible that the contributions of the two parts are larger than if we only consider $U_\textrm{MW}\left(r\right)$ and extend it to the host galaxy \citep{Gao:2015,Wei:2015}. Adopting the Keplerian potential for the Milky way, i.e., $U_{\rm MW}(r)=-GM_{\rm MW}/r$, we obtain the difference of PPN parameters of two photons,
\begin{equation}\label{PPN}
\Delta\gamma\equiv\gamma_1-\gamma_2<\Delta t_\textrm{obs}\left(\frac{GM_\textrm{MW}}{c^3}
\right)^{-1}\ln^{-1}\left(\frac{d}{b}\right),
\end{equation}
where $G=6.68\times 10^{-8}~{\rm erg~cm}~{\rm g}^{-2}$ is the gravitational constant, $M_\textrm{MW}=6\times 10^{11}M_{\odot}$ is the mass of the Milky Way, $d$ is the luminosity distance between source and observer, and $b$ is the impact parameter of the light rays relative to the center of the Milky Way. In this paper, the luminosity distance $d$ is calculated in the concordance $\Lambda$CDM model with cosmological parameters $H_0=70\textrm{ km s}^{-1}\textrm{Mpc}^{-1}$, $\Omega_\textrm{M}=0.3$ and $\Omega_{\Lambda}=0.7$. The impact parameter $b$ can be estimated as
\begin{equation}\label{impact parameter}
b=r_\textrm{G}\sqrt{1-\left(\sin\delta_\textrm{S}\sin\delta_\textrm{G}+\cos\delta_\textrm{S} \cos\delta_\textrm{G}\cos\left(\beta_\textrm{S}-\beta_\textrm{G}\right)\right)^2},
\end{equation}
where $r_\textrm{G}=8.3$ kpc is the distance from the Sun to the galaxy center, $\beta_\textrm{S}$ and $\delta_\textrm{S}$ are respectively the right ascension and declination of the source in the equatorial coordinates, and $\left(\beta_\textrm{G}=17^\textrm{h}45^\textrm{m}40.04^\textrm{s},\delta_\textrm{G}=-29^\circ00'28.1''\right)$ are the coordinates of the Galaxy center \citep{Gillessen:2008qv}.

\section{Constraint From Short GRBs}

Thanks to the launch of various space satellites, such as Compton, Swift, and Fermi, some properties of GRBs have been well researched. Through the analysis of light curves, one can obtain the duration of a GRB, which is usually characterized by $T_{90}$. During $T_{90}$ from 5\% to 95\% of the total photon events in a specific energy band are detected. The observed durations span about 6 orders of magnitude, from milliseconds to thousands of seconds. According to the fact that the distribution of durations is bimodal and separated at about 2 s, \citet{Kouveliotou:1993} proposed a GRBs classification: long bursts with $T_{90}>2$ s and short bursts with $T_{90}<2$ s. These two classes of GRBs are often thought to be produced by two different progenitors. Most of the GRBs spectra can be well fitted with the Band function \citep{Band:1993eg}, which peaks at around a few hundred keV. In some bright GRBs, photons with energy higher than 100 MeV and even tens of GeV have been observed \citep{Abdo:2009,Abdo:2009pg,Ackermann:2010us,Ackermann:2011bu}. X-ray emission is usually weak, and a few emissions are below 10 keV \citep{Piran:1999}.

In this paper, we test EEP using short GRBs detected by the Burst Alert Telescope (BAT) onboard the Swift satellite. BAT can catch photons in the energy band of 15--150 keV, and record the light curves of GRBs. We choose GRBs from the Swift data archive, and only short GRBs with $T_{90}<2$ s are selected. 20 GRBs are finally picked out with measured redshifts in the range $z\in\left[0.093,2.609\right]$. The main properties of the GRB sample are listed in Table~\ref{tab:results}.
\begin{table}
	\centering
	\caption{The 20 short Swift/BAT GRBs. RA and DEC are respectively the right ascension and declination of GRBs in J2000, $z$ is the redshift, $T_{90}$ is the GRB duration in $15-150$ keV energy band, and $\Delta\gamma$ is the difference of PPN parameters between 15 keV and 150 keV.}
	\label{tab:results}
	\begin{tabular}{lccccc} 
		\hline
		GRBs &  RA&  Dec & $z$  & $T_{90}$ [s] & $\Delta\gamma$\\
		\hline
150120A	&$00^h41^m19.2^s$&$+33^\circ58'48.0''$&0.460	&	1.2	&	$3.18\times10^{-8}$\\
150101B	&$12^h32^m10.6^s$&$-10^\circ57'21.6''$&0.093	&	0.018	&	$5.59\times10^{-10}$ \\
141212A	&$02^h36^m40.1^s$&$+18^\circ09'46.8''$&0.596	&	0.3	&	$7.62\times10^{-9}$	\\
140903A	&$15^h52^m05.0^s$&$+27^\circ36'28.8''$&0.351	&	0.3	&	$8.15\times10^{-9}$	\\
140622A	&$21^h08^m36.7^s$&$-14^\circ24'43.2''$&0.959	&	0.13	&	$3.18\times10^{-9}$	\\
131004A	&$19^h44^m25.9^s$&$-02^\circ57'07.2''$&0.717	&	1.54	&	$3.81\times10^{-8}$	\\
130603B	&$11^h28^m53.3^s$&$+17^\circ03'46.8''$&0.356	&	0.18	&	$4.92\times10^{-9}$	\\
101219A	&$04^h58^m20.6^s$&$-02^\circ31'37.2''$&	0.718	&	0.6	&	$1.47\times10^{-8}$	\\
100724A	&$12^h58^m16.6^s$&$-11^\circ05'42.0''$&	1.288	&	1.4	&	$3.39\times10^{-8}$	\\
090510	&$22^h14^m12.5^s$&$-26^\circ35'52.8''$&	0.903	&	0.3	&	$7.75\times10^{-9}$	\\
090426	&$12^h36^m19.7^s$&$+32^\circ58'40.8''$&	2.609	&	1.2	&	$2.75\times10^{-8}$	\\
071227	&$03^h52^m31.7^s$&$-55^\circ57'32.4''$&	0.383	&	1.8	&	$4.90\times10^{-8}$  \\
070724A	&$01^h51^m17.8^s$&$-18^\circ36'36.0''$&	0.457	&	0.4	&	$1.07\times10^{-8}$	\\
070429B	&$21^h52^m01.4^s$&$-38^\circ51'25.2''$&	0.904	&	0.47	&	$1.16\times10^{-8}$	\\
061217	&$10^h41^m36.7^s$&$-21^\circ09'07.2''$&	0.827	&	0.21	&	$5.32\times10^{-9}$	\\
061201	&$22^h08^m19.0^s$&$-74^\circ34'04.8''$&	0.111	&	0.76	&	$2.29\times10^{-8}$	\\
060502B	&$18^h35^m42.5^s$&$+52^\circ37'04.8''$&	0.287	&	0.131	&	$3.67\times10^{-9}$	\\
051221A	&$21^h54^m51.6^s$&$+16^\circ53'16.8''$&	0.547	&	1.4	&	$3.67\times10^{-8}$ \\
050813	&$16^h08^m00.2^s$&$+11^\circ14'38.4''$&	1.800	&	0.45	&	$1.04\times10^{-8}$ \\
050509B	&$12^h36^m11.0^s$&$+28^\circ58'26.4''$&	0.225	&	0.073	&	$2.09\times10^{-9}$ \\
		\hline
	\end{tabular}
\end{table}
In this table, we list the GRB name, the GRB position in the equatorial coordinates (RA and DEC), the redshift $z$ and the duration $T_{90}$. Conservatively speaking, the observed time delay of any two photons in the specific energy band couldn't be longer than the duration. Using $T_{90}$ as the upper limit of $\Delta t_\textrm{obs}$ in equation~(\ref{PPN}), we finally have
\begin{equation}\label{DeltaGamma}
\Delta\gamma<T_{90}\left(\frac{GM_\textrm{MW}}{c^3}
\right)^{-1}\ln^{-1}\left(\frac{d}{b}\right).
\end{equation}
This will give a strict constraint on the difference of PPN parameters in  $15-150$ keV energy band.

The constraints on $\Delta\gamma (15-150~{\rm keV})$ from 20 GRBs are listed in the last column of Table~\ref{tab:results}. Among them eleven GRBs constrain $\Delta\gamma$ at the order of $10^{-8}$, eight GRBs at the order of $10^{-9}$. The strictest limit is given by GRB 150101B, i.e., $\Delta\gamma<5.59\times10^{-10}$. This is because GRB 150101B has the shortest duration among the sample, although the redshift (so the distance) of this GRB is smaller than others. The duration $T_{90}$ makes more contribution to $\Delta\gamma$ than redshift $z$, because $\Delta\gamma$ is proportional to $T_{90}$, while it is inversely proportional to the logarithm of distance. Therefore, to further constrain $\Delta\gamma$, short-duration GRBs rather than high-redshift GRBs are necessary. The constraint on $\Delta\gamma$ from GRB 150101B is about two orders of magnitude smaller than that previously constrained from GRB 090510 ($|\gamma_\textrm{GeV}-\gamma_\textrm{MeV}|<2\times10^{-8}$) \citep{Gao:2015}, and improves the constraint from FRB 150418 ($\Delta\gamma<1-2\times10^{-9}$) \citep{Tingay:2016tgf} by a factor of two.

\section{Discussions and Conclusions}

The validity of EEP has been extensively tested using photons from energetic cosmic transients.
Recently, \citet{Gao:2015} proposed that the time delays between correlated photons from GRBs can be used to constrain EEP. They considered the gravitational field of the Milky Way and constrained the differences of $\gamma$ of photons with energies between MeV and GeV and between eV and MeV. In the observation of GRB 090510, a single 31 GeV photon was detected 0.83 s after the trigger of MeV photons, which coincides in time with the main emission of MeV photons. The time delay between the single 31 GeV photon and MeV photons was used to constrain $\Delta\gamma$, resulting $|\gamma_\textrm{GeV}-\gamma_\textrm{MeV}|<2\times10^{-8}$. This result is not statistically significant, because only one GeV photon was used. We couldn't know whether the GeV photon is emitted simultaneously with MeV photons. We even couldn't ensure if they are emitted in the same region. The other burst used is GRB 080319B. The most conservative time delay between MeV photons and optical photons is 5 s according to the optical-$\gamma$-ray correlation function, which leads to the result $|\gamma_\textrm{eV}-\gamma_\textrm{MeV}|<1.2\times10^{-7}$. This constraint, although statistically significant, is about three orders of magnitudes looser than our results. Recently, \citet{Tingay:2016tgf} used the spectral lag of radio photons in $\sim$GHz energy band from the first localized FRB 150418 and obtained a tighter constraint $\Delta\gamma<1-2\times10^{-9}$, which is still looser than our result by a factor of two.

In this paper, we used the duration of GRB as the upper limit of the observed time delay, and improve previous results by $1\sim 2$ orders of magnitude. Our results are statistically significant. There are hundreds of thousands of photons in the energy band of 15 -- 150 keV, and the energy spectra are continuous. Any two of those photons can be used to constrain the accuracy of EEP. Conservatively speaking, the observed time delay of any two photons in the energy band of 15 -- 150 keV is smaller than the duration $T_{90}$. Thus, using keV photons to constrain EEP is more statistically reliable. The main uncertainty comes from the  intrinsic time delay $\Delta t_{\rm int}$. We assumed that the high energy photons are emitted later (at least not earlier) than low energy photons. This assumption is based on the observational evidence that in some GRBs high energy photons have spectral lag with respect to low energy photons. If $\Delta t_{\rm int}<0$, and the intrinsic time delay exactly cancels with the time delay induced by EEP violation, then the observed time delay will be small. In this case, the constraint on $\Delta\gamma$ may be much looser. However, the probability of such a coincidence happening in all the 20 GRBs is negligible.

In summary, the observed time delay between photons with different energies can be used to test the validity of EEP by constraining the difference of PPN parameters. We investigated the continuous spectra in the energy band $15 - 150$ keV of 20 short GRBs detected by Swift/BAT and used the durations as the most conservative estimation of time delay to constrain EEP. The strictest constraint on the PPN parameter, $\Delta\gamma<5.59\times10^{-10}$, is given by the shortest-duration burst GRB 150101B. This constraint is $1\sim 2$ orders of magnitude tighter than previous results. Moreover, the statistical significance is highly improved.

\section*{Acknowledgements}

We are grateful to X. Li, P. Wang and S. Wang for useful discussions. This work has been funded by the National Natural Science Foundation of China under grant No. 11375203.




\bibliographystyle{mnras}
\bibliography{myreference} 








\bsp	
\label{lastpage}
\end{document}